%% file: Lestrade_dd_nika2_jun2019_12nov2019.tex
\documentclass{webofc}
\usepackage[varg]{txfonts}   
\begin{document}
\title{Debris disks around stars in the NIKA2 era}
%
%

\author{\firstname{J.-F.} \lastname{Lestrade} \inst{\ref{LERMA}}\fnsep\thanks{\email{jean-francois.lestrade@obpm.fr}}
\and \firstname{J.-C.} \lastname{Augereau} \inst{\ref{IPAG}}
\and \firstname{M.} \lastname{Booth} \inst{\ref{JENA}}
\and \firstname{R.} \lastname{Adam} \inst{\ref{LLR},\ref{CEFCA}}
\and  \firstname{P.} \lastname{Ade} \inst{\ref{Cardiff}}
\and  \firstname{P.} \lastname{Andr\'e} \inst{\ref{CEA1}}
\and  \firstname{A.} \lastname{Andrianasolo} \inst{\ref{IPAG}}
\and  \firstname{H.} \lastname{Aussel} \inst{\ref{CEA1}}
\and  \firstname{A.} \lastname{Beelen} \inst{\ref{IAS}}
\and  \firstname{A.} \lastname{Beno\^it} \inst{\ref{Neel}}
\and  \firstname{A.} \lastname{Bideaud} \inst{\ref{Neel}}
\and  \firstname{O.} \lastname{Bourrion} \inst{\ref{LPSC}}
\and  \firstname{M.} \lastname{Calvo} \inst{\ref{Neel}}
\and  \firstname{A.} \lastname{Catalano} \inst{\ref{LPSC}}
\and  \firstname{B.} \lastname{Comis} \inst{\ref{LPSC}}
\and  \firstname{M.} \lastname{De~Petris} \inst{\ref{Roma}}
\and  \firstname{F.-X.} \lastname{D\'esert} \inst{\ref{IPAG}}
\and  \firstname{S.} \lastname{Doyle} \inst{\ref{Cardiff}}
\and  \firstname{E.~F.~C.} \lastname{Driessen} \inst{\ref{IRAMF}}
\and  \firstname{A.} \lastname{Gomez} \inst{\ref{CAB}}
\and  \firstname{J.} \lastname{Goupy} \inst{\ref{Neel}}
\and  \firstname{W.} \lastname{Holland} \inst{\ref{UKATC}}
\and  \firstname{F.} \lastname{K\'eruzor\'e} \inst{\ref{LPSC}}
\and  \firstname{C.} \lastname{Kramer} \inst{\ref{IRAME}}
\and  \firstname{B.} \lastname{Ladjelate} \inst{\ref{IRAME}}
\and  \firstname{G.} \lastname{Lagache} \inst{\ref{LAM}}
\and  \firstname{S.} \lastname{Leclercq} \inst{\ref{IRAMF}}
\and \firstname{C.} \lastname{Lef\`evre} \inst{\ref{IRAMF}}
\and  \firstname{J.F.} \lastname{Mac\'ias-P\'erez} \inst{\ref{LPSC}}
\and  \firstname{P.} \lastname{Mauskopf} \inst{\ref{Cardiff},\ref{Arizona}}
\and \firstname{F.} \lastname{Mayet} \inst{\ref{LPSC}}
\and  \firstname{A.} \lastname{Monfardini} \inst{\ref{Neel}}
\and  \firstname{L.} \lastname{Perotto} \inst{\ref{LPSC}}
\and  \firstname{G.} \lastname{Pisano} \inst{\ref{Cardiff}}
\and  \firstname{N.} \lastname{Ponthieu} \inst{\ref{IPAG}}
\and  \firstname{V.} \lastname{Rev\'eret} \inst{\ref{CEA1}}
\and  \firstname{A.} \lastname{Ritacco} \inst{\ref{IRAME}}
\and  \firstname{C.} \lastname{Romero} \inst{\ref{IRAMF}}
\and  \firstname{H.} \lastname{Roussel} \inst{\ref{IAP}}
\and  \firstname{F.} \lastname{Ruppin} \inst{\ref{MIT}}
\and  \firstname{K.} \lastname{Schuster} \inst{\ref{IRAMF}}
\and  \firstname{S.} \lastname{Shu} \inst{\ref{IRAMF}}
\and  \firstname{A.} \lastname{Sievers} \inst{\ref{IRAME}}
\and  \firstname{P.} \lastname{Th\'ebault} \inst{\ref{LESIA}}
\and  \firstname{C.} \lastname{Tucker} \inst{\ref{Cardiff}}
\and  \firstname{R.} \lastname{Zylka} \inst{\ref{IRAMF}}}

\institute{
\label{LERMA} LERMA, Observatoire de Paris, PSL Research University,CNRS, Sorbonne Universit\'es, UPMC Univ. Paris 06, 75014 Paris, France
\and \label{IPAG} Univ. Grenoble Alpes, CNRS, IPAG, 38000 Grenoble, France     
\and \label{IRAMF} Institut de RadioAstronomie Millim\'etrique (IRAM), Grenoble, France
\and \label{JENA} Astrophysikalisches Institut und Universit\"atssternwarte, D-07745 Jena, Germany
\and \label{UKATC} UK Astronomy Technology Centre, Royal Observatory, Edinburgh, UK
\and \label{LLR} LLR (Laboratoire Leprince-Ringuet), CNRS, \'Ecole Polytechnique, Institut Polytechnique de Paris, Palaiseau, France
\and \label{Cardiff} Astronomy Instrumentation Group, University of Cardiff, UK
\and \label{LPSC} Univ. Grenoble Alpes, CNRS, Grenoble INP, LPSC-IN2P3, 53, avenue des Martyrs, 38000 Grenoble, France
\and \label{CEFCA} Centro de Estudios de F\'isica del Cosmos de Arag\'on (CEFCA), Plaza San Juan, 1, planta 2, E-44001, Teruel, Spain 
\and \label{CEA1} AIM, CEA, CNRS, Universit\'e Paris-Saclay, Universit\'e Paris Diderot, Sorbonne Paris Cit\'e, 91191 Gif-sur-Yvette, France     
\and \label{IAS} Institut d'Astrophysique Spatiale (IAS), CNRS and Universit\'e Paris Sud, Orsay, France    
\and \label{Neel} Institut N\'eel, CNRS and Universit\'e Grenoble Alpes, France
\and \label{Roma} Dipartimento di Fisica, Sapienza Universit\`a di Roma, Piazzale Aldo Moro 5, I-00185 Roma, Italy       
\and \label{IRAME} Instituto de Radioastronom\'ia Milim\'etrica (IRAM), Granada, Spain 
\and \label{CAB} Centro de Astrobiolog\'ia (CSIC-INTA), Torrej\'on de Ardoz, 28850 Madrid, Spain
\and \label{LAM} Aix Marseille Univ, CNRS, CNES, LAM (Laboratoire d'Astrophysique de Marseille), Marseille, France
\and \label{Arizona} School of Earth and Space Exploration and Department of Physics, Arizona State University, Tempe, AZ 85287         
\and \label{IAP} Institut d'Astrophysique de Paris, CNRS (UMR7095), 98 bis boulevard Arago, 75014 Paris, France
\and \label{MIT} Kavli Institute for Astrophysics and Space Research, Massachusetts Institute of Technology, Cambridge, MA 02139, USA
\and \label{LESIA} LESIA, Observatoire de Paris, PSL Research University,CNRS, Sorbonne Universit\'es, UPMC Univ. Paris 06, 75014 Paris, France 
          }
\abstract{%
The new NIKA2 camera at the IRAM 30m radiotelescope was used to observe three known debris disks
in order to  constrain the SED of their dust emission  in the millimeter wavelength domain. 
We have found that the spectral index between the two NIKA2 bands (1mm and 2mm) 
is consistent with the Rayleigh-Jeans regime ($\lambda^{-2}$), unlike the steeper spectra ($\lambda^{-3}$) 
measured in the submillimeter-wavelength domain for two of the three disks $-$ around the stars Vega and HD107146.
We provide a succesful proof of concept to model this spectral inversion in using two populations of dust grains,
those smaller and those larger than a grain radius $a_0$ of 0.5mm. This is obtained in  breaking the slope of the
 size distribution and the functional form of the absorption coefficient of the standard model. 
The third disk $-$ around the star HR8799 $-$  does not exhibit this spectral inversion but is also the youngest.}
\maketitle
\section{Introduction}
\label{intro}
A debris disk orbiting a main sequence star is made of residual planetesimals left over from the  agglomeration processes
during the early phase of planet formation. 
It is akin to the  asteroidal belt and the Kuiper Belt remaining from the planet formation in the solar system.
The largest planetesimals $-$ sub-km to hundreds of km in diameters $-$ around other stars than the Sun 
 have a total cross sectional area  that is insufficient to be detected by our telescopes.
However, when giant planets are present in these distant systems, planetesimals are  gravitationally stirred 
and collide at a significant rate to produce 
a multitude of fragments including a myriad of small dust particles which has a total cross sectional area large enough to be observable 
for stars up to $\sim$ 100 pc. This dust can be detected both in scattered light and in thermal emission as an excess above the stellar 
photospheric level from mid-wave infrared  to millimeter wavelengths or longer. 
This observable dust is a marker of a dynamically active planetesimal belt, indicating that the host star possesses a planetary system.
The discovery of  debris disks was one of the
highlights of the first  infrared  satellite IRAS \cite{Aumann1984}.

Dust emission in excess of photospheric emission can be used to constrain the size distribution of fragments as well as
their composition  \cite{Wyatt2002, Lebreton2012}, and the dynamical state of the associated planetary system \cite{MacGregor2016}.
Theoretically, the equilibrium size distribution of fragments resulting from a collisional cascade is caracterised by the  number density  
$n(a) \propto a^{2-3q_d}$ ($a$ the fragment size) with $q_d=11/6$ for
the infinite self similar collisional model  of asteroids and their debris in the solar system 
for material strength independant of size  \cite{Dohnanyi1969} and \cite{Tanaka1996} \footnote{We recall that  parameter $q_d$ $-$ the power law index 
of the general form for the number density of fragments per unit volume of space and per unit mass $-$ is 
parameter $\alpha$ of eq.~2 in Dohnanyi's theoretical work\cite{Dohnanyi1969}}.
However other size distributions are possible depending on the actual internal strength of the planetesimal  material \cite{Durda1997}, and
on the  dynamical processes at work in the disk \cite{MacGregor2016}.
In addition, noticeably, the stellar radiation pressure blows  the smallest grains out of the system  and significantly perturbates 
the collisional cascade in imprinting a wavy pattern to the size distribution at small scale
\cite{Thebault2003}.

If micronic dust grains, abundant in this paradigm, absorb efficiently stellar radiation, they re-radiate inefficiently  at long wavelengths. This increases 
their temperature and steepens the slope of the SED at long wavelengths. Constraining observationally the SED in this part of the spectrum is key to caracterise the type of
collisional cascade at the root of the dust production in debris disks.

The SONS JCMT/SCUBA2 Legacy survey  detected 48 debris disks in the submillimeter domain ($\lambda=450 \mu$m and $850 \mu$m) 
and found a large range of spectral slopes with $\beta \in [0 - 2.7]$  ($S_{\nu} \propto \lambda^{-(2+\beta)}$) \cite{Panic2013, Holland2017}, 
which correspond to an index $q_d$ possibly larger than 2 for the collisional cascade, {\it i.e.} size distribution steeper than the standard self similar collisinal model.   
However, with a sample of a dozen of debris disks observed 
by SMA and ALMA at 1.3mm and 850$\mu$m,  another group has found the range $q_d \in [1.61  - 1.88]$ \cite{MacGregor2016}, leading to 
an opposite conclusion for the size distribution.
We have started  NIKA2 observations during winter 2017 to complement these studies with photometry at 1.2mm and 2mm.

\section{NIKA2 observations of three stars with known debris disks}
\label{sec-1}

The three stars  HD107146, Vega and HR8799 were selected because their disks offered the largest flux densities at 2mm 
predicted in extrapolating from the SCUBA2 data in \cite{Holland2017}.
Observations of these three stars were conducted  on October 29 and 30  2017   
at the IRAM 30m radiotelescope at Pico de Veleta in Spain in fair weather conditions ($\tau_{225GHz}=0.2-0.30$) 
using the  new NIKA2 camera with its three   
arrays paved with Kinetic Inductance Detectors (KIDs) enclosed in a cryogenic system \cite{Catalano:2014nml,NIKA2-Adam, Calvo2016}. 
Arrays A1 and A3 providing the two linear polarisations at 1.2mm were combined to measure the flux density at this wavelength 
 and array A2 was used to measure the total flux density at 2 mm. 
Uranus and MWC349 were used as primary and secondary calibrators to establish and check the flux density scale  \cite{NIKA2-Commissioning}. 
Data were calibrated and imaged using the NIKA2 data reduction
pipeline. The flux densities were determined by fitting the standard gaussians at 1.2mm and 2mm as used to calibrate 
on Uranus in the pipeline \cite{NIKA2-Commissioning}.

\begin{figure}[h]
\resizebox{14cm}{!}{\includegraphics[scale=0.8]{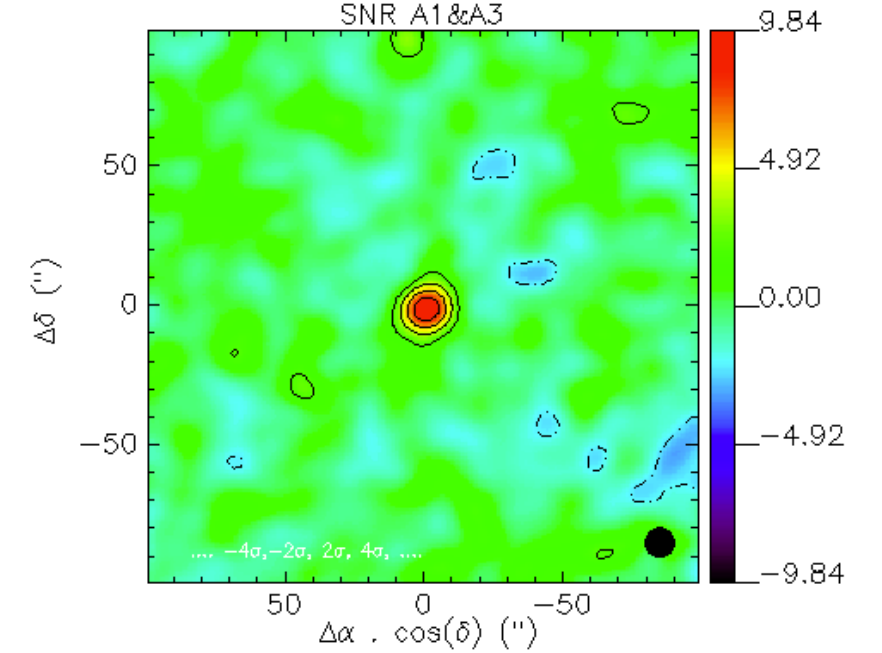}  \includegraphics[scale=0.8]{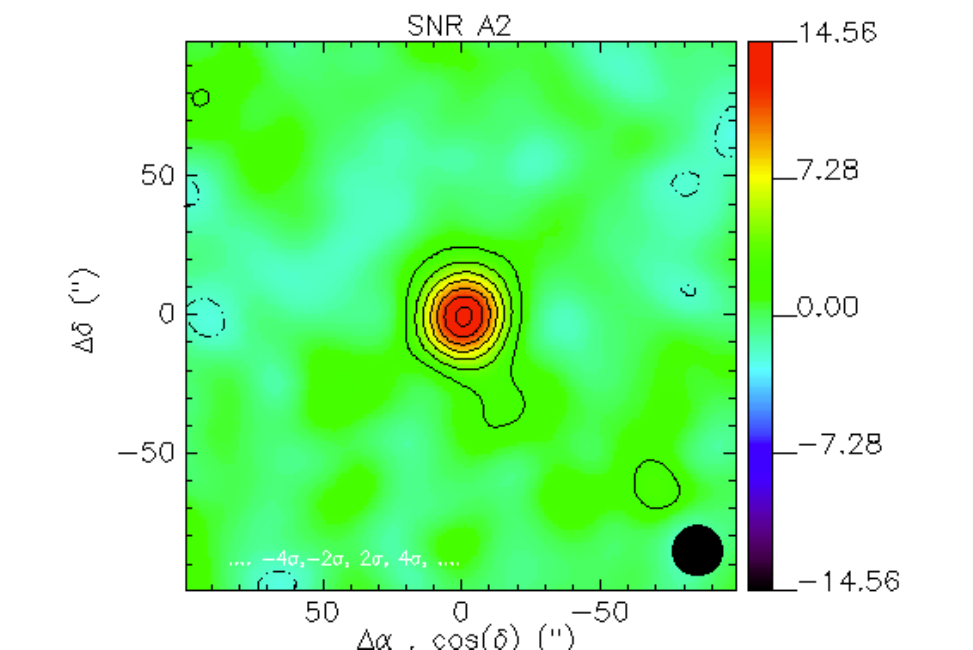}  \includegraphics[scale=0.4,angle=-90,height=70mm,origin=rb]{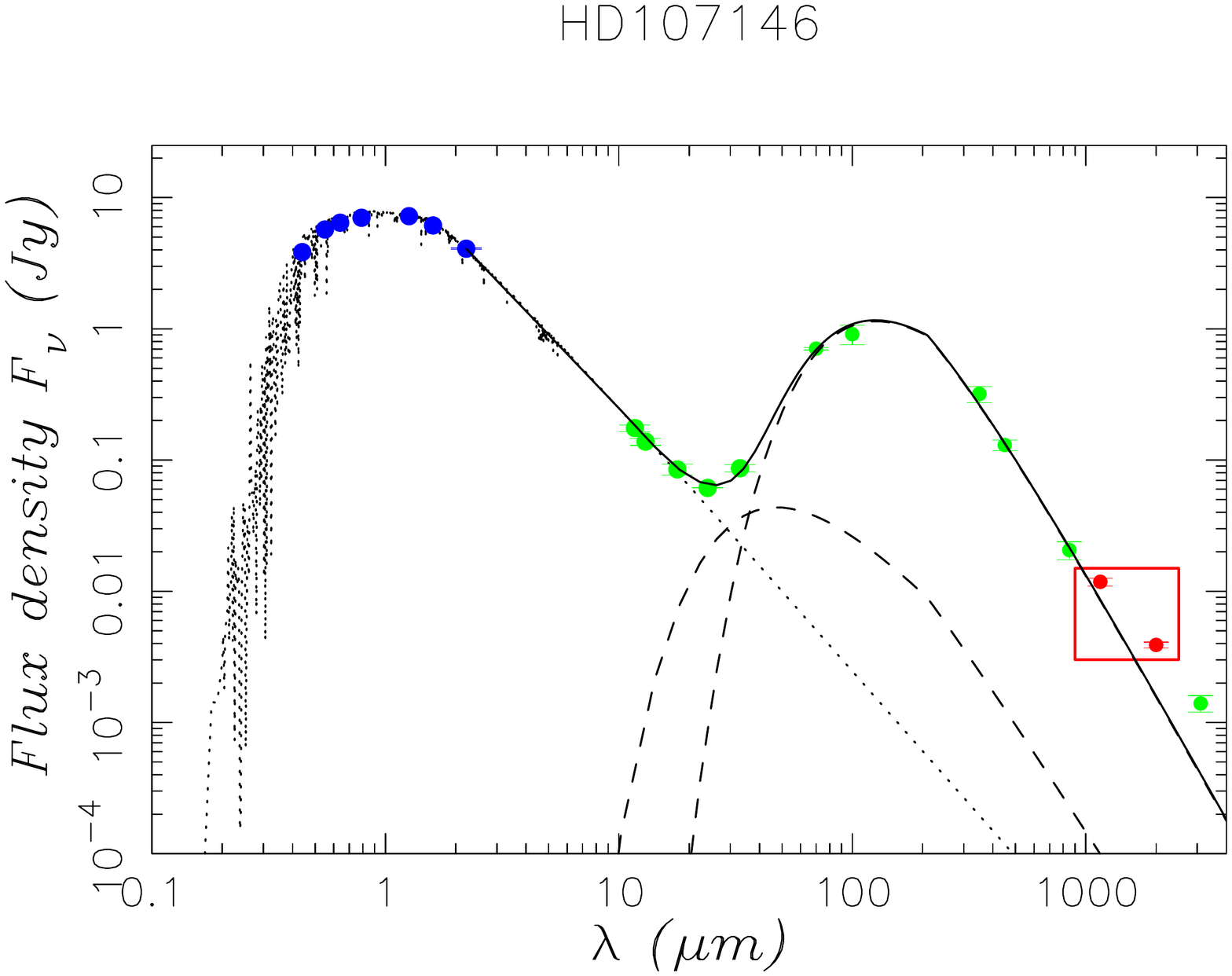}}
\resizebox{14cm}{!}{\includegraphics[scale=0.8]{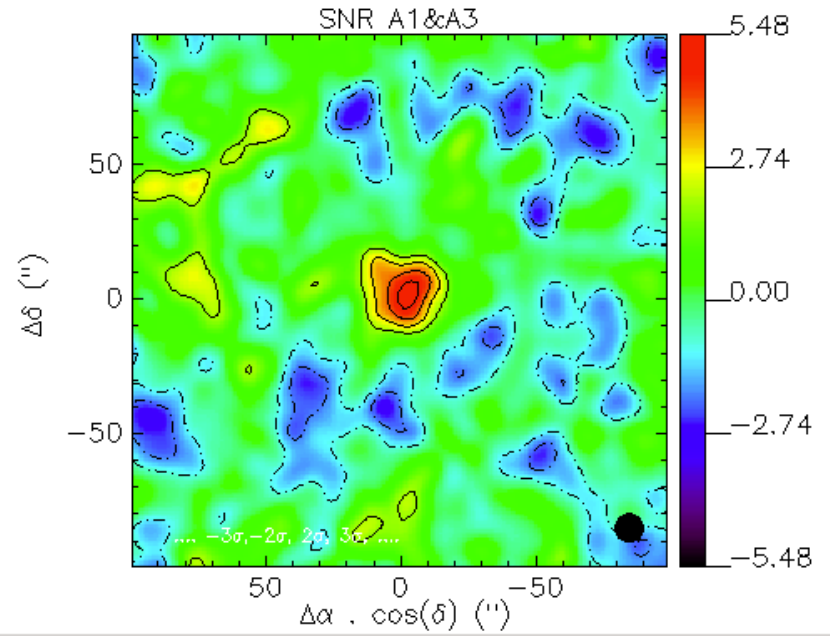}   \includegraphics[scale=0.8]{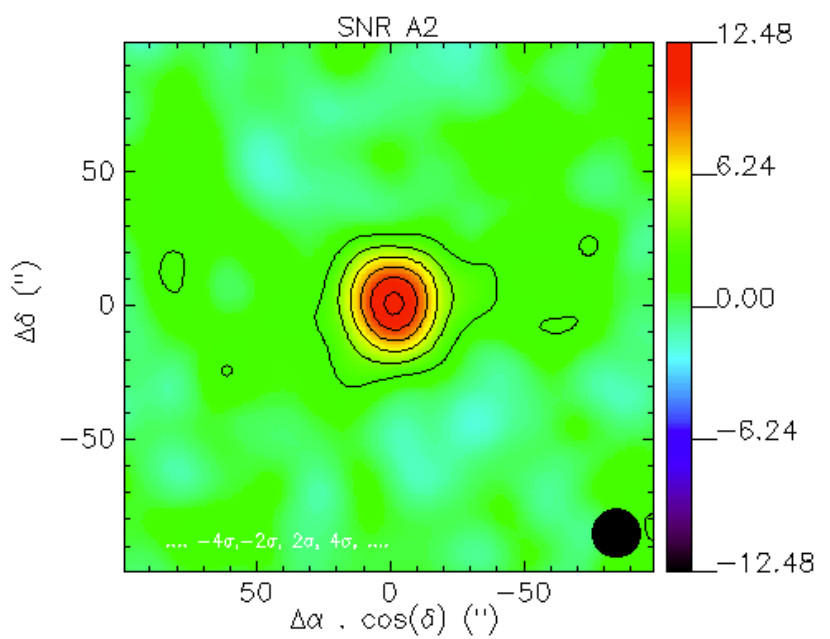}   \includegraphics[scale=0.4,angle=-90,height=70mm,origin=rb]{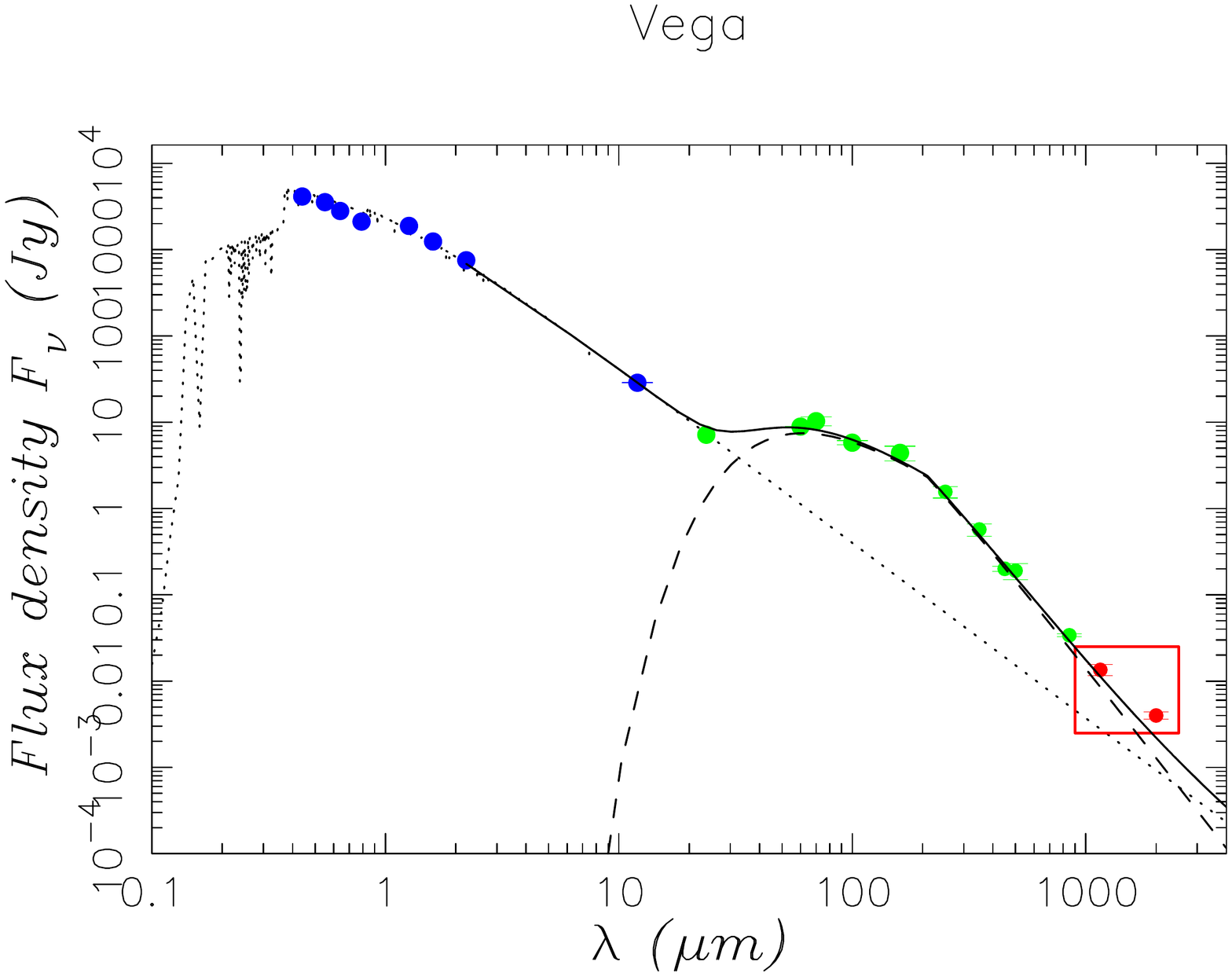} }
\resizebox{14cm}{!}{\includegraphics[scale=0.8]{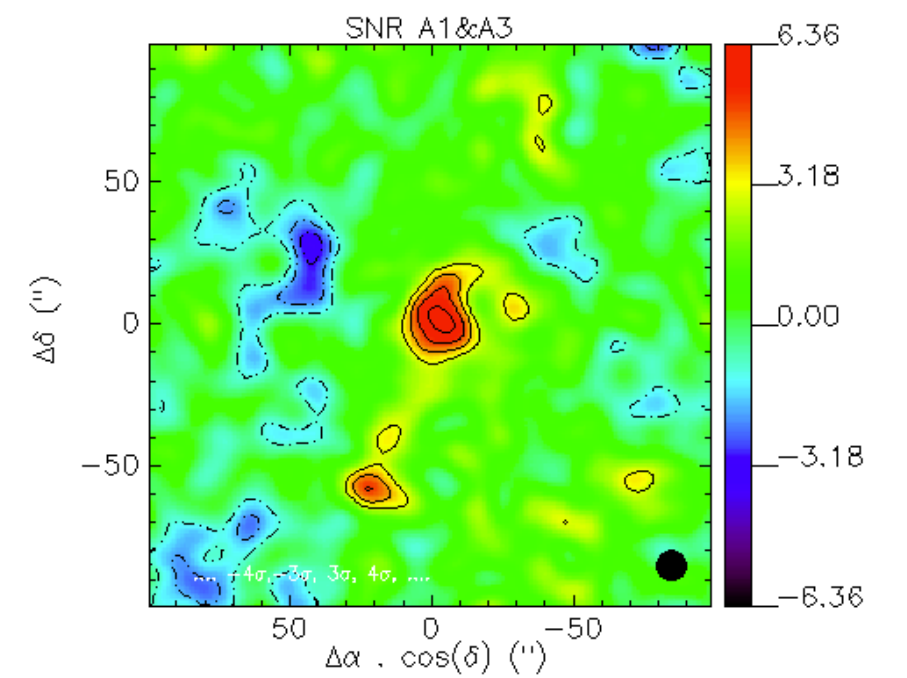} \includegraphics[scale=0.8]{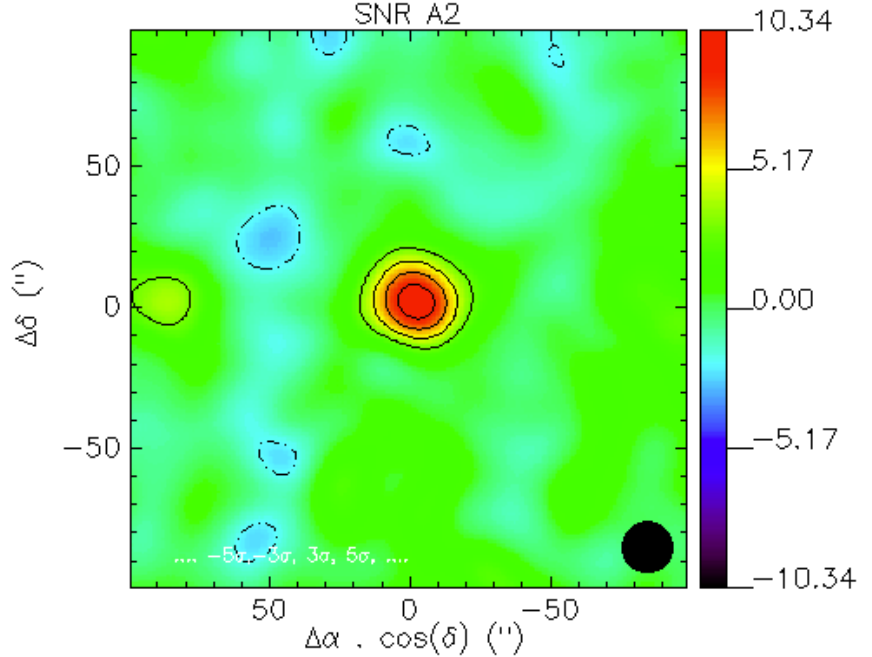} \includegraphics[scale=0.4,angle=-90,height=70mm,origin=rb]{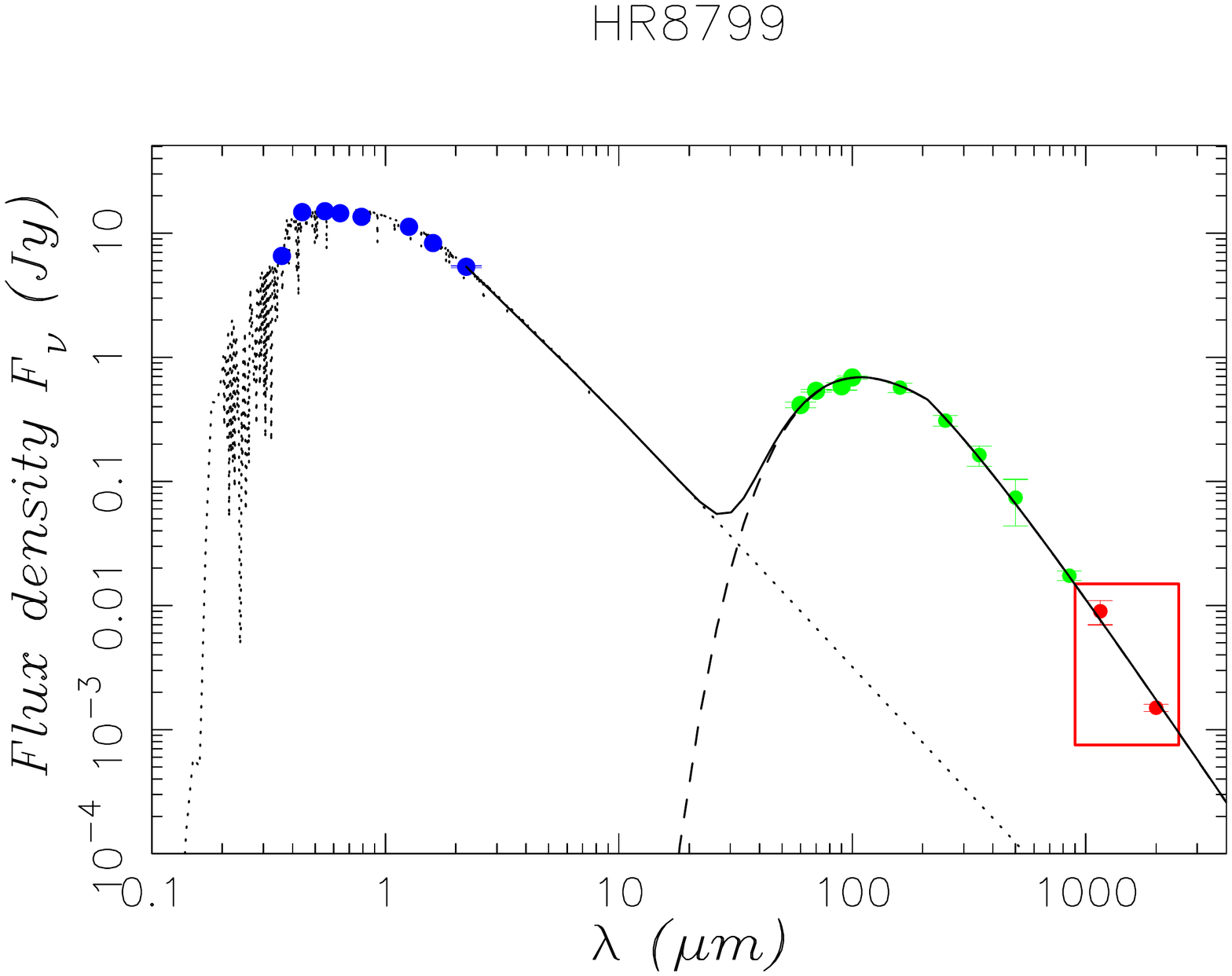}}
\caption{From top to bottom, NIKA2 SNR images of the G2V star HD107146 (age 80-200 Myr) at 28.5pc, the AOV star Vega (age 400-600 Myr) at 7.76 pc, and the A5V star HR8799 (age 20-50 Myr) at 39 pc (from top to bottom). Lower contours are
 $+2\sigma$ for HD107146 and Vega, and $+3\sigma$ for HR8799.  See color scale for SNR. Beams are in black.
 {\bf Left:} 1mm map (arrays A1 and A3 combined), {\bf Middle:} 2mm map (array A2), {\bf Right:} 
SED with the new NIKA2 photometric points at 260 GHz and 150 GHz in red and boxed. The SED of the disk emission has been fitted only 
to the photometric points at $\lambda < 1$mm ; it is apparent on the SEDs that HD107146 and Vega exhibit
a spectral inversion beyond this wavelength while HR8799 does not. Blue dots are optical photometric data to  which the NexGen photospheric model 
is fit. Green dots are the far IR 
to millimeter wavelength data. Red dots are NIKA2. A grey body emissivity is fit to the flux densities at $\lambda=450\mu$m and $\lambda=850\mu$m 
by eye for the dust ($\beta=0.8$) 
and provides a reference model to show excesses at longer wavelengths for HD107146 and Vega. Dashed lines are the dust emission models, in 
the case of HD107146 a two narrow belt model is required to match both the mid IR and millimeter domains.  }
\label{fig:images}   
\end{figure}

Images of the three targets as well as their SED's with the two NIKA2 flux densities 
are plotted  in  Fig.~\ref{fig:images}.  The final values of the NIKA2 flux densities and detailed comparison to the literature                
({\it e.g.} \cite{Hughes2012,Marino2018}) will be published elsewhere. HD107146 is unresolved within uncertainties 
(the main beam FWHMs  are $11''$ and $17.5''$ at 1.2mm and 2mm, respectively). 
We shall carry a detailed comparison with the double ring structure found by ALMA for this disk \cite{Marino2018}. 
Vega and HR8799 are slightly extended at the 30m/NIKA2. The North-West extension of the source HR8977 in Fig.~\ref{fig:images} is also present in the SCUBA2 image 
at 850$\mu$m and ascribed to a background galaxy \cite{Holland2017}.

The spectral indices of the SED measured between 1153$\mu$m  and 2000$\mu$m by NIKA2 are in Table~\ref{tab:beta} as well as values measured between 450$\mu$m 
and 850$\mu$m by SONS \cite{Holland2017} for comparison. For two of the disks $-$ around the stars Vega and HD107146 $-$, we have found
that their NIKA2 spectral indices are consistent with the Rayleigh-Jeans regime ($\lambda^{-2}$), unlike their steeper spectra ($\lambda^{-3}$)
measured at shorter wavelengths by SONS. It is interesting to notice that for HD107146 this spectral inversion 
is also supported by the flux density measured at $\lambda=3.1$mm by the OVRO millimeter interferometer \cite{Carpenter2005} and included in the SED of this star 
in Fig.~\ref{fig:images}. The third debris disk  $-$ around the star HR8799 $-$ does not show this spectral inversion but is also the youngest
in our small sample.

\begin{table}
\centering
\caption{$\beta$ measured by SONS, ALMA and NIKA2  (this work) }
\label{tab:beta}       
\begin{tabular}{lcccc}
\hline
Star     & age (Myr)  & $\beta_{850/1100}^{(1)}$   &  $\beta_{450/850}^{(2)}$     & $\beta_{1153/2000}$  \\
         &            &    (ALMA)           &    (SCUBA2)             &    (NIKA2)       \\\hline
Vega     & 400-600    &                     &    $0.9\pm0.12$       & $\sim 0.0\pm0.32$ \\
HD107146 & 80-200     & $0.61 \pm0.49$      &       0.8-1.0$^{(*)}$     & $\sim 0.0\pm0.18$  \\
HR8799   & 20-50      &                     &   $1.7\pm0.20$       & $1.38\pm0.45$ \\\hline
\multicolumn{5}{l}{\footnotesize 
(1) ALMA see \cite{Marino2018} ; (2) SONS see \cite{Holland2017} ; (*) based on grey body fit of $\beta$ and $\lambda_0$ in \cite{Holland2017}.}
\end{tabular}
\end{table}

\section{Proof of concept to model the spectral inversion}
\label{sec-2}

\begin{figure}[h]
\resizebox{14cm}{!}{\includegraphics[scale=1.0, angle=-90]{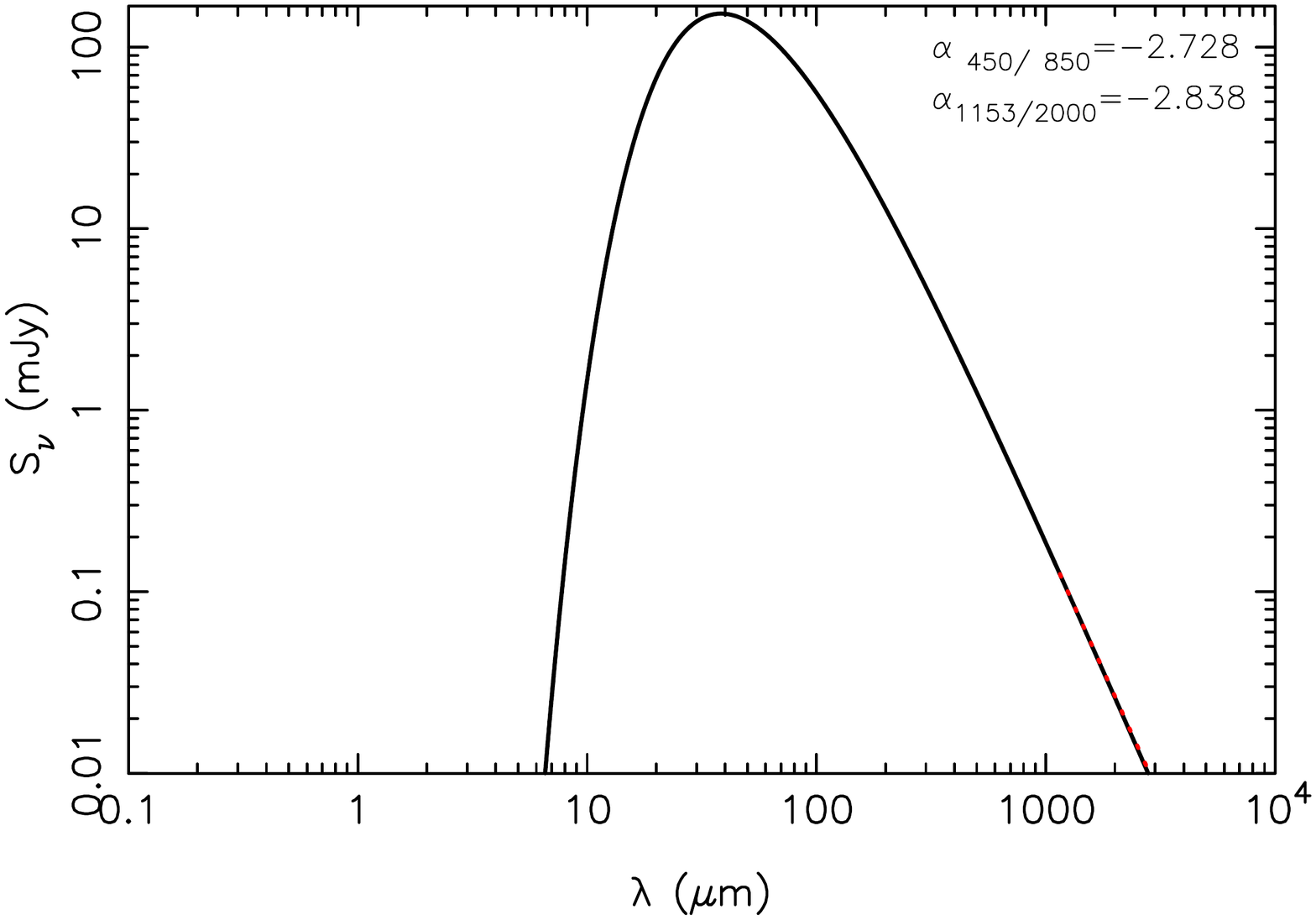}  \includegraphics[scale=1.0, angle=-90]{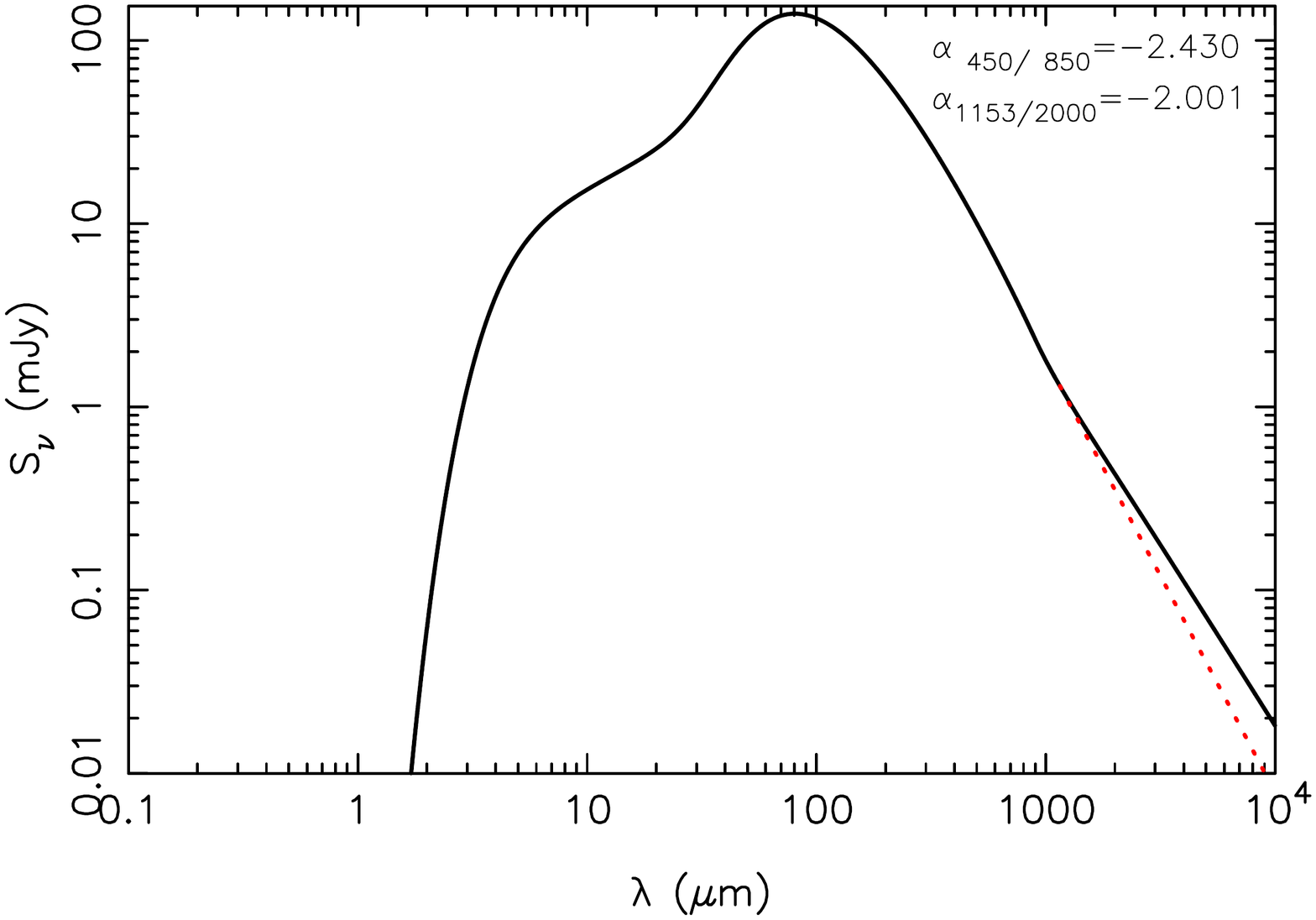}  }
\caption{Proof of concept to test the possibility of a change in the spectral index at millimeter wavelengths in the SED 
of the debris disk around HD107146. {\bf Left :} standard model using a single population of dust grains with the size distribution characterised 
by the index $q_{d}=2$ and a shallow absorption coefficient ($Q_{abs}$ power-law index is 1 here). This yields the spectral 
indices  $\alpha_{450/850}=-2.73$ and $\alpha_{850/1153}=-2.84$, {\it i.e.} no spectral inversion is produced. {\bf Right :} 
model using two populations of grains caracterised by a break in the grain size distribution ($q_{d,1}=1.7$ and $q_{d,2}=2.0$) 
and in the absorption coefficient ($Q_{abs}$ power-law indices are $5$ and 0.0 (black body)) for grains smaller 
and larger than $a_0=0.45$mm, respectively. This latter model does result in a spectral index inversion, 
changing from  $\alpha_{450/850}=-2.43$  to $\alpha_{1153/2000}=-2.00$. 
The dotted red line extrapolates the grey body behavior found in the submillimeter domain into the millimeter domain 
for comparison with the modelled SED. The other parameters of the model are the dust fractional luminosity $f_d=5~10^{-3}$, 
minimum and maximum grain radii : $0.1\mu$m and 10cm, stellar luminosity 1.1 L$_{\odot}$, stellar radius 1.09 R$_{\odot}$, 
disk radius 70AU, and star distance 28.5pc.}
\label{fig:SED}
\end{figure}

As a proof of concept we have developed a model to test whether or not a change in the spectral index is possible at all. First,
  we have modelled  a narrow belt of spherical dust grains and used the standard
size distribution  $dN= N_0 a^{2-3q_d} da$ and 
the schematic absorption coefficient $Q_{abs}(a,\lambda) = 2 a/\lambda$ when $2 a < \lambda$ and $Q_{abs}(a,\lambda) =1$ otherwise.
Dust temperature was computed at thermal equilibrium, making the temperature of micronic-sized grains 
significantly larger than their black body temperature as it is well known. 
An example of such a SED is given in  Fig.~\ref{fig:SED} (left) resulting in the spectral index  $\alpha_{1153/2000} = -(2+\beta) = -2.84$ between the two NIKA2 bands.
  
Then, we have sought to invert the spectral index  from typically $\beta \sim 1$ 
in the submillimeter domain to $\beta \sim 0$ at $\lambda > 1$mm as observed for Vega and HD107146  with NIKA2.
First, we  broke the slope of the dust size distribution $(q_{d,1}, q_{d,2})$ at some grain radius $a_0$ varied 
over the range $[0.1{\rm mm}, 2{\rm mm}]$  and in assuming continuity of the grain number densities 
at $a_0$  $({\it i.e.}~N_1 a_0^{2-3q_{d,1}}=N_2 a_0^{2-3q_{d,2}})$. 
This interval is where there is a major break in the grain size distribution when interaction between stellar radiation pressure 
and small grains at the bottom of the collisional cascade of planetesimals is accounted for  \cite{Thebault2003} and \cite{Kim2018}.
In varying $q_{d,1}$ and $q_{d,2}$ from  1.7 to 2.0, we  could not reproduce the observed  
spectral inversion with this model. So, we also forced a change in the schematic absorption coefficient described above 
in retaining a steep power-law  $Q_{abs}(a,\lambda)$ for all grains of  radii smaller than $0.5$mm 
and in using  $Q_{abs}(a,\lambda)=1$ (black body) at all wavelengths for all larger grains. Although this  abrupt
change of the dust properties at some grain radius $a_0$ is not physically satisfactory and requires a smoother transition, it is interesting to note 
that it does yield a spectral inversion  at $\lambda = 1$mm as seen 
in Fig.~\ref{fig:SED} (right). We note that in this model for HD107146, the temperature of micronic dust grains raises above 200 K producing the 
mid-infrared secondary peak seen in this figure. In our modeling, the fractional 
dust luminosity $f_d$ has been kept constant, so the emitting areas of the two grain populations 
are redistributed in a way that makes the total emitting surface unchanged when $q_{d,1}$ and $q_{d,2}$ are varied.

\section{Concluding remarks}
\label{sec-3}

Further studies are planned to sophisticate the model in including a wavy form for the size distribution of the micronic grains
as established on theoretical ground by \cite{Thebault2003} and in including a more physical absorption coefficient based
on the real dust properties of complex aggregates (shape, porosity, composition, etc) computable with the flexible 
software SIGMA \cite{Lefevre2019}. We also plan to expand the NIKA2 observations to a larger sample of debris disks
to investigate whether or not the black body behavior of dust in the millimeter domain found for Vega and HD107146 
is a general property of debris disks, possibly with the exceptions of the youngest stars as already found for HR8799 
in our small sample. In this case, the less evolved collisional cascade could possibly be less perturbed by 
stellar radiation pressure retaining a more abundant population of small grains as shown in \cite{Thebault2003}.  
A complete study is in progress and will be published.



\section*{Acknowledgements}
J-F Lestrade gratefully acknowleges support of the Programme National de Plan\'etologie (PNP) of CNRS.   
\input{acknowledgements}

%
%
%

\end{document}

%% file: acknowledgements.tex
We would like to thank the IRAM staff for their support during the campaigns. The NIKA dilution cryostat has been designed and built at the Institut N\'eel. In particular, we acknowledge the crucial contribution of the Cryogenics Group, and in particular Gregory Garde, Henri Rodenas, Jean Paul Leggeri, Philippe Camus. This work has been partially funded by the Foundation Nanoscience Grenoble and the LabEx FOCUS ANR-11-LABX-0013. This work is supported by the French National Research Agency under the contracts "MKIDS", "NIKA" and ANR-15-CE31-0017 and in the framework of the "Investissements d’avenir” program (ANR-15-IDEX-02). This work has benefited from the support of the European Research Council Advanced Grant ORISTARS under the European Union's Seventh Framework Programme (Grant Agreement no. 291294). We acknowledge fundings from the ENIGMASS French LabEx (R. A. and F. R.), the CNES post-doctoral fellowship program (R. A.), the CNES doctoral fellowship program (A. R.) and the FOCUS French LabEx doctoral fellowship program (A. R.). R.A. acknowledges support from Spanish Ministerio de Econom\'ia and Competitividad (MINECO) through grant number AYA2015-66211-C2-2.

%% file: Lestrade_dd_nika2_jun2019_12nov2019.bbl
\begin{thebibliography}{}
%
%


\bibitem{Aumann1984}
  H.~H.~Aumann {\it et al.},
 ApJ. \ {\bf   278}, L23 (1984)

\bibitem{Wyatt2002}
 M.C. Wyatt \&  W.R.F. Dent,
 MNRAS, {\bf 334}, 589 (2002)

\bibitem{Lebreton2012}
 J.  Lebreton {\it et al.}, 
 Astron.\ Astrophys.\ {\bf 539}, 17 (2012)

\bibitem{MacGregor2016}
  M.~A. MacGregor {\it et al.},
 ApJ., {\bf 823}, 79 (2016)

\bibitem{Dohnanyi1969}
 J.S. Dohnanyi,
 J. Geophys. Res., {\bf 74}, 2531 (1969) 

\bibitem{Tanaka1996}
 H. Tanaka, S. Inaba, K. Nakazawa, 
 Icarus, {\bf 123}, 450 (1996)

\bibitem{Durda1997}
 D.D. Durda  \&  S.F. Dermott, 
 Icarus, {\bf 130}, 140 (1997)

\bibitem{Thebault2003}
 P. Thebault, J.C. Augereau, H., Beust, 
 Astron.\ Astrophys.\  {\bf 408}, 775 (2003)

\bibitem{Panic2013}
 O. Pani\'c {\it et al.},
 MNRAS, {\bf 435}, 1037 (2013)

\bibitem{Holland2017}
 W. S. Holland {\it et al.},
 MNRAS, {\bf 470}, 3606 (2017)

\bibitem{Catalano:2014nml}
 A.~Catalano {\it et al.},
  Astron.\ Astrophys.\  {\bf 569}, A9 (2014)

\bibitem{NIKA2-Adam}
 R.~Adam {\it et al.},
  Astron.\ Astrophys.\  {\bf 609}, A115 (2018)

\bibitem{Calvo2016}
 M. Calvo {\it et al.}, 
 Journal of Low Temperature Physics, {\bf 184}, 816 (2016)

\bibitem{NIKA2-Commissioning}
 L.~Perotto {\it et al.},
arXiv : 1910.02038 (2019)

\bibitem{Hughes2012}
A.M. Hughes  {\it et al.},
ApJ, 750, 82 (2012)

\bibitem{Marino2018}
S. Marino  {\it et al.},
MNRAS, 479, 5423 (2018)

\bibitem{Carpenter2005}
 J.~M. Carpenter {\it et al.},
AJ., {\bf 129}, 1049 (2005)

\bibitem{Kim2018}
 M. Kim {\it et al.},
  Astron.\ Astrophys.\  {\bf 618}, 38 (2018)

\bibitem{Lefevre2019}
 C.~Lef\`evre {\it et al.},
submitted to  Astron.\ Astrophys. and see this proceedings



\end{thebibliography}
